\documentclass[11pt,twoside]{article}
\usepackage{CAGN2019}
\usepackage{graphicx}

\usepackage[T1]{fontenc} 

\usepackage{latexsym}
\usepackage{verbatim}

\usepackage{ifpdf}  
\ifpdf  
      \DeclareGraphicsExtensions{.pdf,.png,.jpg}  
\else  
      \DeclareGraphicsExtensions{.eps}  
\fi 

\begin{document}

\vskip 1.0cm
\markboth{J. E. M\'endez-Delgado et al.}{Radial distribution of helium in the Milky Way}
\pagestyle{myheadings}
%
%
\vspace*{0.5cm}
\parindent 0pt{Poster}


\vspace*{0.5cm}
\title{Radial distribution of helium in the Milky Way}

\author{J. E. M\'endez-Delgado$^{1,2}$, C. Esteban$^{1,2}$ and J. Garc\'ia-Rojas$^{1,2}$}
\affil{$^1$Instituto de Astrof\'isica de Canarias, E-38200 La Laguna, Spain.\\
$^2$Departamento de Astrof\'isica, Universidad de La Laguna, E-38206, La Laguna, Spain.}

\begin{abstract}

We present preliminary results of the study of the radial distribution of helium in the Milky Way. We use 37 spectra from 23 Galactic HII regions observed with VLT, GTC and Magellan Telescope. Using PyNeb, we calculate the abundance of He$^+$ with all the He$\thinspace$I detected lines. In most cases, both the average abundance of He$^+$ and its associated dispersion are higher when triplet lines are used. Although corrections for collisional processes are considered in the calculations, the differences in the He$^+$ abundance are not negligible between both spin configurations. This suggests that self-absorption processes are important in most triplet lines. Four ICFs were used to estimate the contribution of He$^0$ in the total abundance. The resulting radial distribution of helium has a negative slope when singlet lines are used, regardless of the ICF used. 


\end{abstract}

\section{Introduction}

The star formation history and the gas flows determine the distribution of chemical abundances in galaxies. The existence of radial gradients of chemical abundances in spiral galaxies is a well known fact, and is thought to be caused by 'inside-out' galaxy formation scenario, where the number of generations of stars that have existed is smaller as the galactocentric distance increases.\\

Almost no metals were created in the beginning of the Universe. Thus, the difference in metallic content between two stellar generations is noticeable and a radial gradient can be traced. On the other hand, the biggest fraction of helium was created during the Primordial Nucleosynthesis and the determination of its radial gradient is highly sensitive to errors and uncertainties in the calculation of He/H ratio.\\


To determine a radial gradient in the Galactic distribution of helium requires a sample of objects with good quality spectra that covers a wide range of Galactocentric distances. HII regions are suitable for this purpose since they trace the present-day chemical composition of the Galaxy.

\section{Procedure}
\label{procedure}

\begin{figure}[t]  
\begin{center}
\hspace{0.25cm}
\includegraphics[height=5.0cm]{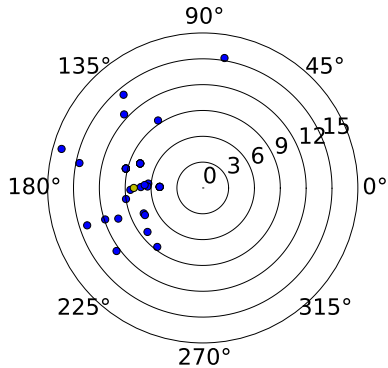}
\caption{Distribution of HII regions of the sample. The Galactocentric distance is presented in kpc. The yellow dot indicates the position of the Sun (8 kpc).}
\label{hiimap}
\end{center}
\end{figure}

Our sample consists on 37 deep spectra of 23 Galactic HII regions, covering a range of radial distance between 5 and 17 kpc. In Figure \ref{hiimap} the distribution of the HII regions in the Galaxy disk is shown.\\ 

The data have been obtained with the high-resolution spectrograph UVES at VLT, OSIRIS at GTC and MagE at Magellan Telescope \citep{gr_04_NGC3576, ce_04_ORION, gr_05_S311, gr_06, gr_07, ce_13_NGC2579, ce_16, ce_17}. The spectral resolution and wavelength coverage allowed us to detect several He$\thinspace$I recombination lines in each spectrum in both spin configurations: triplet and singlet. \\


We calculated the abundance of He$^+$ using PyNeb \citep{pyneb_ref} with all the detected He$\thinspace$I lines in the sample. Some lines as $\lambda$3889, $\lambda$7065, $\lambda$7281 and $\lambda$9464 were discarded from the analysis for being deeply affected by line blending, self-absorption, collisions or blending with sky lines. Some other lines were also discarded in particular cases, as $\lambda$5016 in the Orion Nebula \citep{ce_04_ORION}. The effective recombination coefficients used were those calculated by \cite{porter_2013}, which include corrections for collisional effects. \\ 

To calculate the total helium abundance, the He$^0$/H$^+$ ratio was estimated using four Ionization Correction Factors (ICF) based on similarities between ionisation potentials of sulfur and/or oxygen and helium \citep{peimbert_77, kunth_83, peimbert_92, zhang_03}. Using Galactocentric distance estimates from the literature for each region of the sample, we derived the radial distribution of helium, analysing separately the results based on singlet and triplet lines.


\section{Results and conclusions }
\label{results}

\begin{figure}[h]  
\begin{center}
\hspace{0.25cm}
\includegraphics[height=4.5cm]{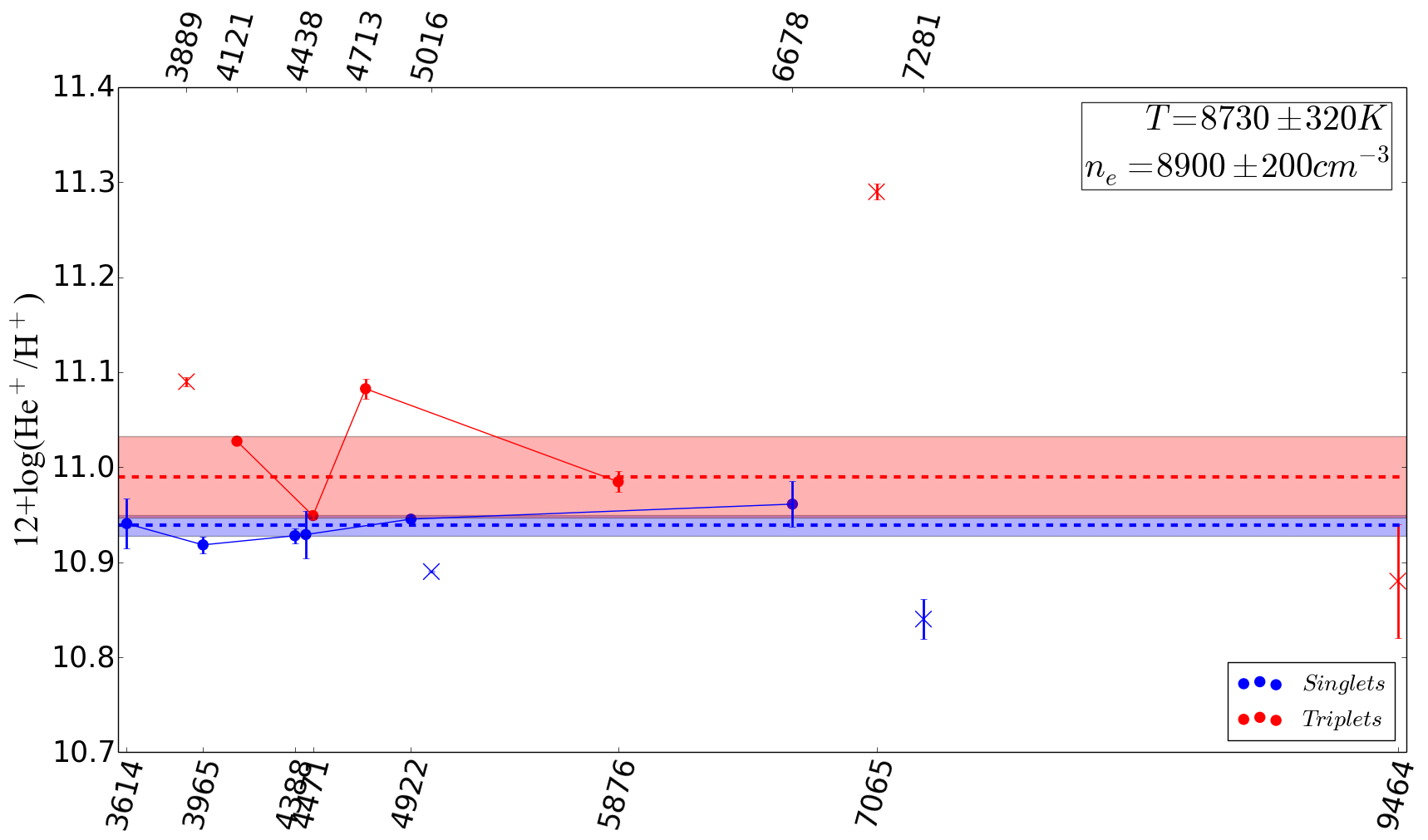}
\caption{Abundance of He$^+$ obtained with several recombination lines. Horizontal lines represent the average value for the singlets and triplets. The coloured band indicates the associated uncertainty. The red symbols represent triplets while blue ones represent singlets. Crosses indicate discarded lines.}
\label{orion}
\end{center}
\end{figure}

\begin{figure}[h!]  
\begin{center}
\hspace{0.25cm}
\includegraphics[height=4.5cm]{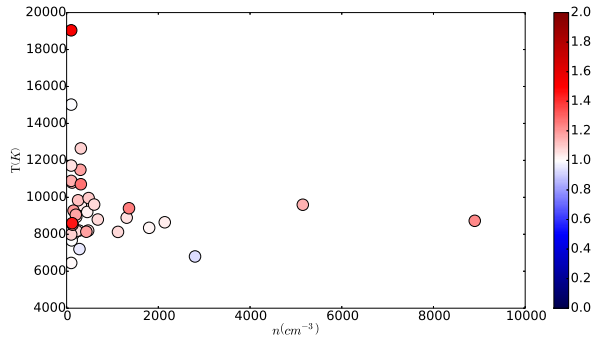}
\caption{Representation of the He$^{+}$/H$^+$ ratio determined for He$\thinspace$I 5876\AA$\thinspace$ with respect to the mean He$^+$/H$^+$ ratio determined from singlets (indicated by the color of the circles) as a function of T$_e$ and $n_e$.}
\label{lineas}
\end{center}
\end{figure}

Estimates of He$^+$/H$^+$ and their associated dispersions were comparatively higher using triplet lines.  As an example, Figure \ref{orion} shows calculations of He$^+$/H$^+$ obtained with every detected He$\thinspace$I line in the Orion Nebula. Figure \ref{lineas} shows the overestimate of He$^+$/H$^+$ in calculations based on the triplet He$\thinspace$I $\lambda$5876 line compared to the average value from singlets.\\

\begin{figure}[h]  
\begin{center}
\hspace{0.25cm}
\includegraphics[height=4.5cm]{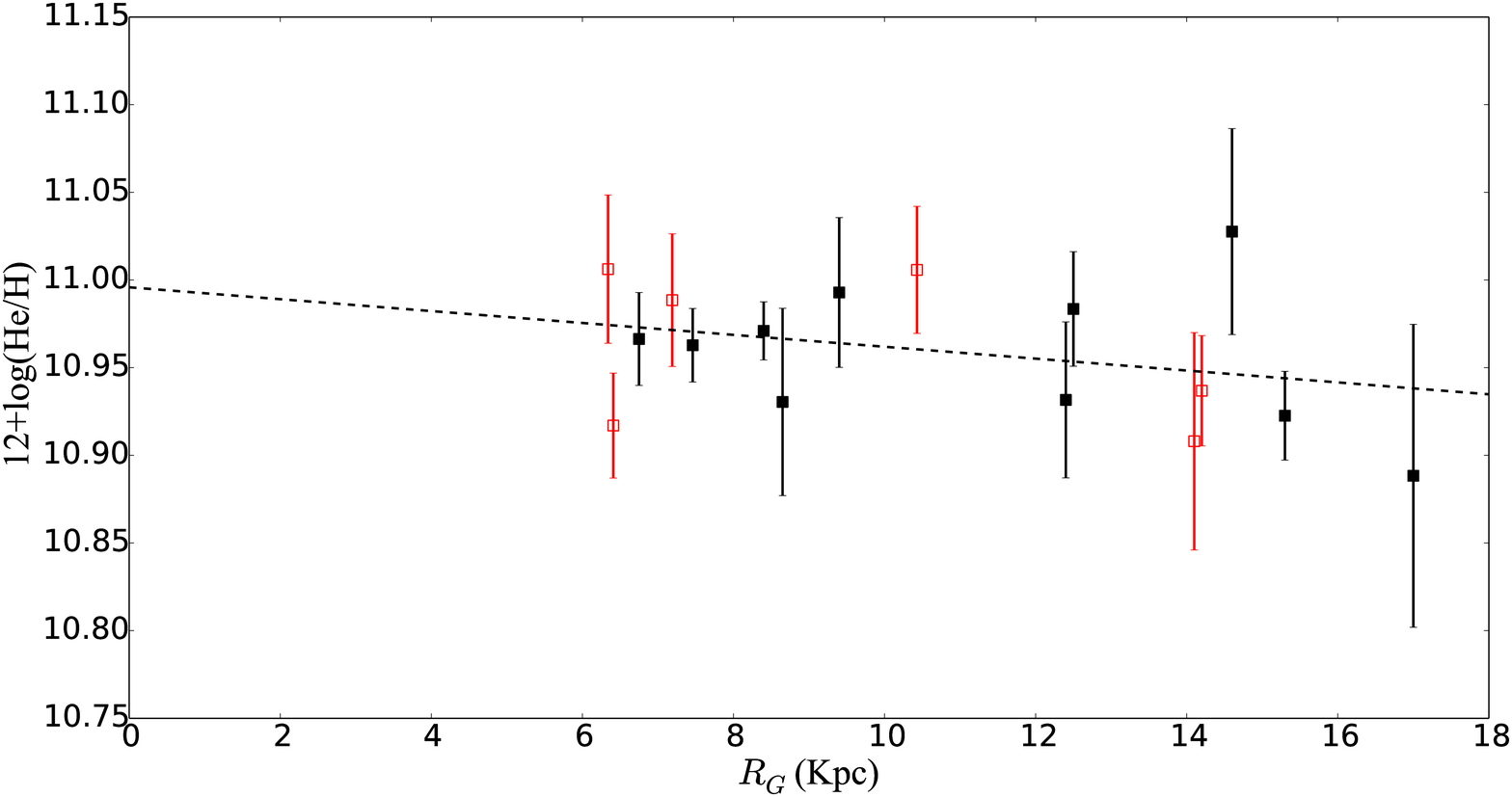}
\caption{Radial distribution of He/H calculated with singlet lines and the average value of all ICF used. Red symbols have ICF$\geq$1.2 and were not
used for the linear fit.}
\label{radial}
\end{center}
\end{figure}

The resulting radial distribution of helium has negative slope when singlet lines were used regardless of the ICF used. In the case of triplet lines, the slope depends on the ICF used. Calculations based on triplet lines are affected by self-absorption mechanisms that are not easy to correct. Figure \ref{radial} shows the radial distribution of helium based on singlet lines and the average value if all the ICF are considered. 

Our preliminar results are summarised as follows:
\begin{enumerate}
\item{Almost all triplet lines comparatively provide overabundances in the He$^{+}$/H$^{+}$ ratio. The dispersion is also larger. Corrections for collisional processes are considered in the He abundance determinations. This suggests that the effects of self-absorption are not negligible in most triplet lines.}
\item{Discarding triplet lines, the radial distribution of helium has a negative slope regardless of the ICF used. Nevertheless, due to the uncertainty introduced by the ICFs, errors in the slope are still consistent with a flat distribution.}
\item{HII regions associated with Wolf-Rayet or evolved O stars present helium overabundances.}
\end{enumerate}

\acknowledgments CEL and JRG acknowledge support from the project AYA2015-65205-P. JGR acknowledges support from an Advanced Fellowship from the Severo Ochoa excellence program (SEV-2015-0548).

\bibliographystyle{aaabib}
\bibliography{examplebib}

\end{document}